\documentclass{article}
\usepackage{spconf,amsmath,graphicx}
\usepackage{tabularx}
\usepackage{adjustbox}
\usepackage{booktabs}
\usepackage{multirow}
\usepackage[table,xcdraw]{xcolor}
\usepackage{hyperref}
\usepackage{indentfirst}
\title{Gammatonegram Representation for End-to-End Dysarthric Speech Processing Tasks: Speech Recognition, Speaker Identification, and Intelligibility Assessment}
\name{Aref Farhadipour\textsuperscript{1}, Hadi Veisi\textsuperscript{2}}
\address{1 Department of Computational Linguistics, University of Zurich, Zurich, Switzerland.\\2 Faculty of New Sciences and Technologies, University of Tehran, Tehran, Iran. \\ \texttt{\{areffarhadi\}@gmail.com;\{h.veisi\}@ut.ac.ir}}
\begin{document}
%\ninept
%
\maketitle
\begin{abstract}
Dysarthria is a disability that causes a disturbance in the human speech system and reduces the quality and intelligibility of a person's speech. Because of this effect, the normal speech processing systems cannot work correctly on this impaired speech. This disability is usually associated with physical disabilities. Therefore, designing a system that can perform some tasks by receiving voice commands in the smart home can be a significant achievement. In this work, we introduce Gammatonegram as an effective method to represent audio files with discriminative details, which can be used as input for convolutional neural networks. In other words, we convert each speech file into an image and propose an image recognition system to classify speech in different scenarios. The proposed convolutional neural networks are based on the transfer learning method on the pre-trained Alexnet. This research evaluates the efficiency of the proposed system for speech recognition, speaker identification, and intelligibility assessment tasks. According to the results on the UASpeech dataset, the proposed speech recognition system achieved a 91.29\% word recognition rate in speaker-dependent mode, the speaker identification system acquired an 87.74\% recognition rate in text-dependent mode, and the intelligibility assessment system achieved a 96.47\% recognition rate in two-class mode. Finally, we propose a multi-network speech recognition system that works fully automatically. This system is located in a cascade arrangement with the two-class intelligibility assessment system, and the output of this system activates each one of the speech recognition networks. This architecture achieves a word recognition rate of 92.3\%.
\end{abstract}
\section{Introduction}
\label{sec:intro}
Speech is the act of conveying emotions and thoughts through vocal sounds to engage in communication with others. However, certain factors, such as illness or physical disability, can result in speech taking on an unintelligible form, thereby hindering the communication process. Individuals who suffer from dysarthria cannot produce natural speech due to limited control over the articulatory aspects of their brain. Furthermore, these individuals often face physical disabilities that impede their ability to perform simple daily tasks effectively.\\
\indent Artificial Intelligence (AI)-based systems have the potential to assist humans in various ways, and aiding individuals with disabilities has always been a prominent area of focus. AI systems can provide a consistent and predefined level of performance, unaffected by environmental or mental factors, when individuals cannot perform specific tasks for various reasons. For individuals with speech disorders, having a system that can automatically process their speech to enhance their quality of life is highly advantageous. For instance, in smart home scenarios designed for disabled individuals, basic tasks such as operating the television, controlling lighting fixtures, and interacting with computers can be made more accessible through Automatic Speech Recognition (ASR) systems. These ASR systems can receive and recognize voice commands, allowing disabled individuals to interact with their environment effectively.\\
\indent However, designing an ASR system that correctly performs for impaired and highly variable speech poses a significant challenge. Typical ASR systems developed for normal speech may not perform well when applied to impaired speech \cite{mengistu2011comparing}. Therefore, it is necessary to develop specific ASR systems tailored to impaired speech, capable of learning the unique characteristics of such speech and delivering acceptable performance.\\
\indent In recent years, deep learning has shown remarkable advancements in various signal processing domains \cite{park2022review,hinton2012deep}. Two-dimensional Convolutional Neural Networks (CNNs) have played a crucial role in image processing \cite{krizhevsky2012imagenet}. However, researchers have explored the same strategy for one-dimensional CNNs in speech processing \cite{yang2021decentralizing}. As an innovation, this study proposes a two-dimensional CNN to develop the systems for three scenarios: ASR, speaker identification, and intelligibility assessment. Additionally, we introduce a cascade multi-network ASR system based on the intelligibility levels of speakers. This system aims to enhance the ASR system's overall performance by leveraging speakers' intelligibility information. We used the UA-speech dataset for dysarthric individuals \cite{kim2008dysarthric} and employed transfer learning to train the networks, particularly in scenarios with limited data availability \cite{zhang2021waste}.\\
\indent Traditionally, speech processing systems have relied on short-term speech features, which are inefficient for dysarthric speech \cite{qian2023survey}. However, we offer a different approach by considering the overall view of an audio file. Our system makes decisions based on a general representation of a voice command, considering these characteristics of dysarthric speech. This is because dysarthric speech often exhibits interruptions in the middle of words, particularly in explosive phonemes and repeated syllables in a periodic manner. The duration of these events can vary depending on the individual's mental and physical conditions. Therefore, analyzing the speech at the word level or considering high-level features can be beneficial.\\
\indent To this end, we proposed the Gammatonegram representation, a weighted version of the traditional spectrogram. Human speech has a particular characteristic where most information is concentrated in the low-frequency range from 50 to 5000 Hz \cite{kent2008acoustic}. The Gammatone filter-bank operates non-linearly for low and high frequencies, providing high resolution for low frequencies and low resolution for high frequencies. This behavior makes Gammatonegrams an efficient representation of speech. Using the Gammatongram image to represent dysarthric speech files is one of our innovations. The experiment results demonstrated that CNNs can perform better for different speech processing scenarios when we used Gammatonegrams as input.\\
\indent The remainder of the article is organized as follows: Section \ref{sec:related} analyzes the related works in dysarthric speech processing. Section \ref{sec:method} explains the methodology that yields the objective of this research. Section \ref{sec:result} reports the system parameters and experimental results. Comparison with the previous works is reported in Section \ref{sec:comp}, and Section \ref{sec:conc} presents the discussion and conclusions.\\

\section{Related Works }
\label{sec:related}
\indent This study contains several systems in three ASR, speaker identification, and intelligibility assessment tasks. This subsection reports some of the related works in these categories.\\
\indent Dysarthric speech recognition is one of the most interesting tasks in impaired speech processing. Most conventional dysarthric speech recognition systems used Hidden Markov Models (HMMs) with several states to model the sequential structure of the speech signal and Gaussian Mixture Models (GMMs) to model the distribution of the features in each state \cite{young2002htk}. \\
\indent In recent years, impaired speech processing performances have grown thanks to the development of deep neural network (DNN) algorithms. Kim et al. \cite{kim2018dysarthric} adopted convolutional long short-term memory recurrent neural networks to model dysarthric speech in a speaker-independent situation. Authors in \cite{liu2019use} attempted to use a gated neural network to explore the robust integration of pitch features to improve disordered speech recognition performance. The study in \cite{bhat2018dysarthric} proposed a denoising autoencoder to enhance dysarthric speech and improve feature extraction. Shahamiri \cite{shahamiri2021speech} proposed a speech vision system for dysarthria speech recognition. It generated synthetic voicegrams for all words and speakers. This method delivered an average word recognition rate of 64.71\%. Some works focused on applying meta-learning to find an end-to-end model initialization for dysarthric speech recognition \cite{wang2021improved}. This paper introduced a base model pre-trained from large-scale normal speech data and proposed methods to meta-update the base model by incorporating across-dysarthric speakers' knowledge into the re-initialized model. Speaker adaptation results on the UASpeech dataset achieved a 54.2\% relative word recognition rate.\\
\indent In \cite{liu2021recent}, a set of novel modeling techniques were employed, including neural architectural search, data augmentation model-based speaker adaptation, and cross-domain generation of visual features within an audio-visual speech recognition system framework. Combining these techniques produced a word error rate of 25.21\% on the UA Speech dataset. The multi-stream model introduced in \cite{yue2022raw} consists of convolutional and recurrent layers. It allows for fusing the vocal tract and excitation components. Moreover, they proposed a system with various features, studied the training dynamics, explored the usefulness of the data augmentation, and provided interpretation for the learned convolutional filters. Their best model reaches 40.6\% and 11.8\% word error rates for dysarthric and typical speech, respectively. Takashima et al., \cite{takashima2019end} acquired an end-to-end ASR framework trained by not only the speech data of a Japanese person with an articulation disorder but also the speech data of a physically unimpaired Japanese person and a non-Japanese person with an articulation disorder to relieve the lack of training data of a target speaker.\\
\indent In \cite{shahamiri2023dysarthric}, a customized deep transformer architecture has been proposed. To deal with the data scarcity problem, a two-phase transfer learning pipeline was designed to leverage healthy speech, investigate neural freezing configurations, and utilize audio data augmentation, and in the best situation, a word recognition rate of 67\% has been reported. Almadhor et al. \cite{almadhor2023e2e} proposed a spatio-temporal dysarthric ASR system using a spatial CNN and multi-head attention transformer to extract the speech features visually. Their system utilized transfer learning to generate synthetic leverage and visuals, resulting in a recognition rate of 20.72\% for the UA-Speech database. Yu et al. \cite{yu2023multi} proposed a Multi-stage Audio Visual-HuBERT framework by fusing the dysarthric speech's visual and acoustic information. They offered to use the AV-HuBERT framework to pre-train the recognition architecture of fusing audio and visual information of dysarthric speech. The knowledge gained by the pre-trained model was applied to address the over-fitting problem of the model. The best word error rate of the proposed method was 13.5\% on moderate dysarthric speech. In \cite{rathod2023transfer} a transfer learning approach using the Whisper model was utilized to develop a dysarthric ASR system. Using the Whisper-based method, a word recognition average rate of 59.78\% was obtained for UA-Speech Corpus, based on the Bi-LSTM classifier model.\\
\indent Few studies have been published on dysarthric speaker recognition tasks. One of our previous works \cite{farhadipour2018dysarthric} described the performance of the typical ANN-based system with deep belief network-based features. This system was implemented in single and multi-network modes. In the single-network and text-independent mode, the best results on the UA speech dataset were yielded with 80.1\% speaker identification accuracy for 16 dysarthric speakers. In another work, \cite{kadi2016fully} presented a new approach to improve the analysis and classification of disordered speech. For this purpose, an ear model was introduced. This ear model provided relevant auditory-based cues combined with the usual Mel-Frequency Cepstral Coefficients (MFCC) to represent atypical speech utterances. The experiments were carried out using data from Nemours and Torgo databases of dysarthric speech. gaussian mixture models, support vector machines, and hybrid systems were tested and compared in the context of dysarthric speaker identification. The experimental results achieved a correct speaker identification rate of 97.2\%. However, the challenge of data scarcity was not addressed, which is the concern of the proposed system of our work.\\
\indent Salim et al. \cite{salim2023constant} evaluated the performance of the automatic speaker verification system by comparing Constant-Q Cepstral Coefficients (CQCC) and MFCC features and their combination. The study involved training separate i-vector and x-vector models using MFCC and CQCC features alone and in combination and improved the i-vector and x-vector model’s equal error rates by 15.07\% and 22.75\%, respectively. In \cite{salim2022automatic}, the x-vector models were trained and compared using individual MFCC, prosodic variables, and combinations. The proposed system achieved an 87.34\% recognition rate.\\
\indent Some researchers have worked on speech intelligibility assessment or severity level measurement. In \cite{gupta2021residual}, a new technique to detect dysarthric severity levels was proposed. The authors presented time-domain, frequency-domain, and Teager energy operator analysis of dysarthric speech to justify spectrogram as a feature representation particularly capable of capturing unstructured spectral energy density distributions. Quantifying dysarthria severity based on a residual neural network with short speech segments was reported 98.9\% recognition rate on the UA speech dataset.\\
\indent Al-Qatab et al. \cite{al2021classification} examined the acoustic features and feature selection methods to improve the classification of dysarthric speech. Four acoustic features, including prosody, spectral, cepstral, and voice quality, were used for feature extraction. Furthermore, six classification algorithms were evaluated. The best classification accuracy was 95.80\%. A comparative study on the classification of dysarthria severity levels using different deep learning techniques and speech-disorder specific features computed from prosody, articulation, phonation, and glottal functioning were evaluated on DNN models \cite{joshy2022automated}. In the best situation, the proposed system gave an accuracy of 93.97\% under the speaker-dependent scenario and 49.22\% under the speaker-independent scenario for the UA-Speech database. Hall et al. in \cite{hall2023investigation} reported the optimal setup of deep learning–based dysarthric intelligibility assessment and explained different evaluation strategies. Results indicate an average of 78.2\% classification accuracy for unforeseen low intelligibility speakers, 40.6\% for moderate intelligibility speakers, and 40.4\% for high intelligibility speakers.\\
\indent In \cite{nikhil2023few} a few-shot approach using a transformer model was employed. This whisper-large-v2 transformer model trained on a subset of the UASpeech dataset containing medium intelligibility level patients achieved an accuracy of 85\%. Moreover, the multiclass model achieved an accuracy of 67\%.  Venugopalan et al., \cite{venugopalan2023speech} developed dysarthric speech intelligibility classifiers on 551,176 disordered speech samples contributed by a diverse set of 468 speakers, with a range of self-reported speaking disorders and rated for their overall intelligibility on a five-point scale.\\
\indent Based on the previous research, it has been observed that the current systems and algorithms, although highly efficient for normal speech, still face significant challenges regarding dysarthric speech. These systems need to undergo further development and refinement. One domain that can enhance the efficiency of such systems is feature extraction. Particularly, it is advisable to focus on high-level features due to the substantial variations in dysarthric speech. Additionally, image processing systems have shown promise in addressing these challenges. Hence, this study proposes using Gammatonegram representation as features and a two-dimensional CNN to improve the performance of dysarthric speech processing. Moreover, we evaluate the proposed methodology in all three tasks.\\
\indent Furthermore, we have discovered that implementing a multi-network scenario can significantly benefit individuals with dysarthric speech. This is because dysarthric speech exhibits a wide range of severity with a corresponding diversity in speech characters. Consequently, it is more effective to train individual networks for each class of intelligibility. Since some of the previous works proposed multi-network ASR systems, they all need a human as an assistant to activate the corresponding sub-network based on users' speech intelligibility level. To create a fully automated multi-network scenario, it is essential to assign speech files to their corresponding sub-network automatically. To this end, we have proposed a cascade architecture based on the intelligibility assessment system to feed the multi-network ASR system.\\
 \begin{figure*}[ht]
  \includegraphics[width=\textwidth]{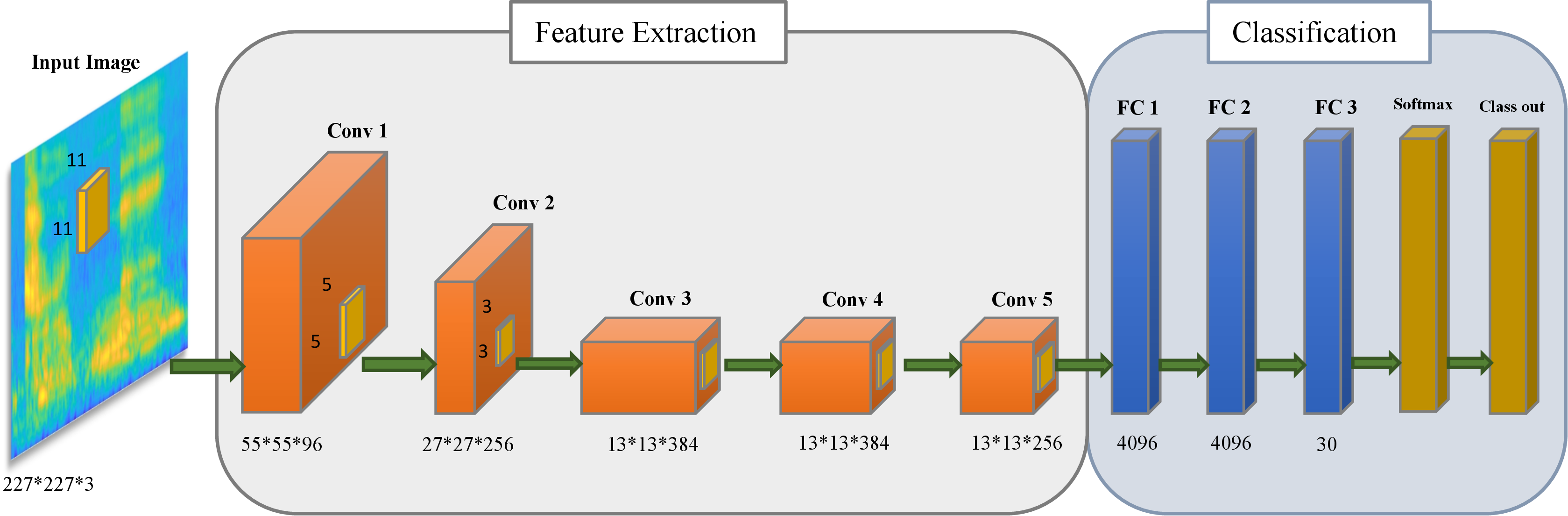}
%  \vspace{2.0cm}

%\end{minipage}

\caption{Diagram of the architecture of Alexnet with feature extraction and classification parts}
\label{fig:f1}
\end{figure*}
\section{Methodology}
\label{sec:method}
\indent This section presents the methods and algorithms utilized in this study, including the description of transfer learning, introduction of Gammatonegram, UA dysarthria speech dataset, and presentation of the utilized Voiced Activity Detector (VAD) algorithm.\\
\subsection{Transfer Learning}
\label{ssec:TL}
\indent CNNs are widely used algorithms in image processing. The term "convolutional" refers to the fact that these networks consist of one or more layers that utilize the convolution operator. Typically, a CNN is composed of two main parts. The first part is responsible for feature extraction and processing of input information through convolutional layers. During the learning process, this part of the network learns to understand visual patterns by employing convolutional multilayer processing. The second part of the network is a classifier that utilizes the features extracted in the first part to construct a model for each class. The network can associate a given speech file with the appropriate class based on the extracted features.\\
\begin{figure*}
  \includegraphics[width=\textwidth]{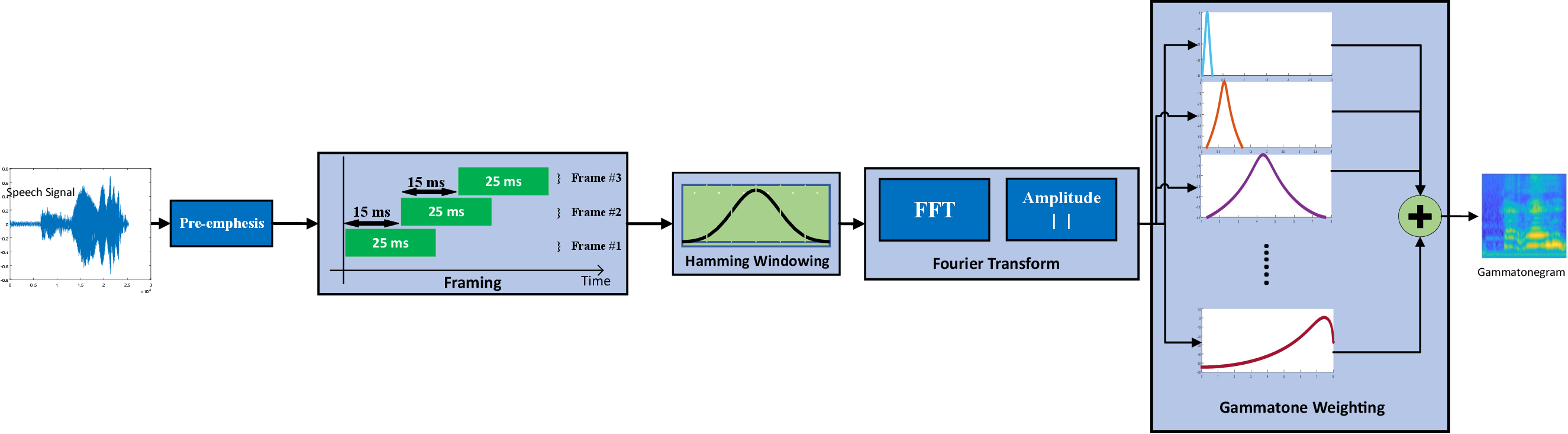}
%  \vspace{2.0cm}

%\end{minipage}

\caption{Block diagram of Gammatonegram extraction steps}
\label{fig:f2}
\end{figure*}
\indent CNNs typically require a large amount of training data to give optimal performance. However, pre-trained CNNs can be modified and reused in limited-data scenarios. These pre-trained models contain information about the input data's dimensions and content. The model's parameters are predetermined in this situation, including the number and type of layers, architecture, and layer connections. Transfer learning is a technique that leverages the weights and parameters of a pre-trained CNN for a new task. Transfer learning eliminates the need for extensive training data by utilizing the knowledge gained from previous training. This is particularly advantageous in low-data conditions as it allows the network to have a pre-existing understanding of vision.\\
\indent The Alexnet is a classic CNN with five convolutional layers to extract more valuable features in deeper layers \cite{krizhevsky2012imagenet}. The last convolutional layer connects to three fully connected layers. The outputs of these layers use the ReLU activation function. The last layers are the softmax and classifier, which determine the output based on the 1000 pre-trained classes. The input of this network is a colored image with dimensions of 227*227*3. The architecture of this network includes about 60 million parameters and more than 650,000 neurons. This network was trained with more than one million images from the Imagenet dataset \cite{deng2009imagenet}. Therefore, according to the classical structure of this network, we used it as the primary network for transfer learning. The structure and parameters of the Alexnet are shown in Fig. \ref{fig:f1}. To create a network for our tasks, we use the feature extraction part of Alexnet and replace new fully connected, softmax, and classifier layers in the classification part to learn the new classes.\\
\indent The study utilizes Gammatonegrams as visual representations of audio signals for input into the CNN. A Gammatonegram is an image that depicts the amplitude or energy of speech signals at different frequency bands and their time of occurrence \cite{pour2014Gammatonegram}. This allows the CNN to process the audio information in a format suitable for image-based analysis.
\subsection{Gammatonegram}
\label{ssec:gamma}
\begin{figure}[htb]

%\begin{minipage}[b]{1.0\linewidth}
  \centering
  \centerline{\includegraphics[width=8.5cm]{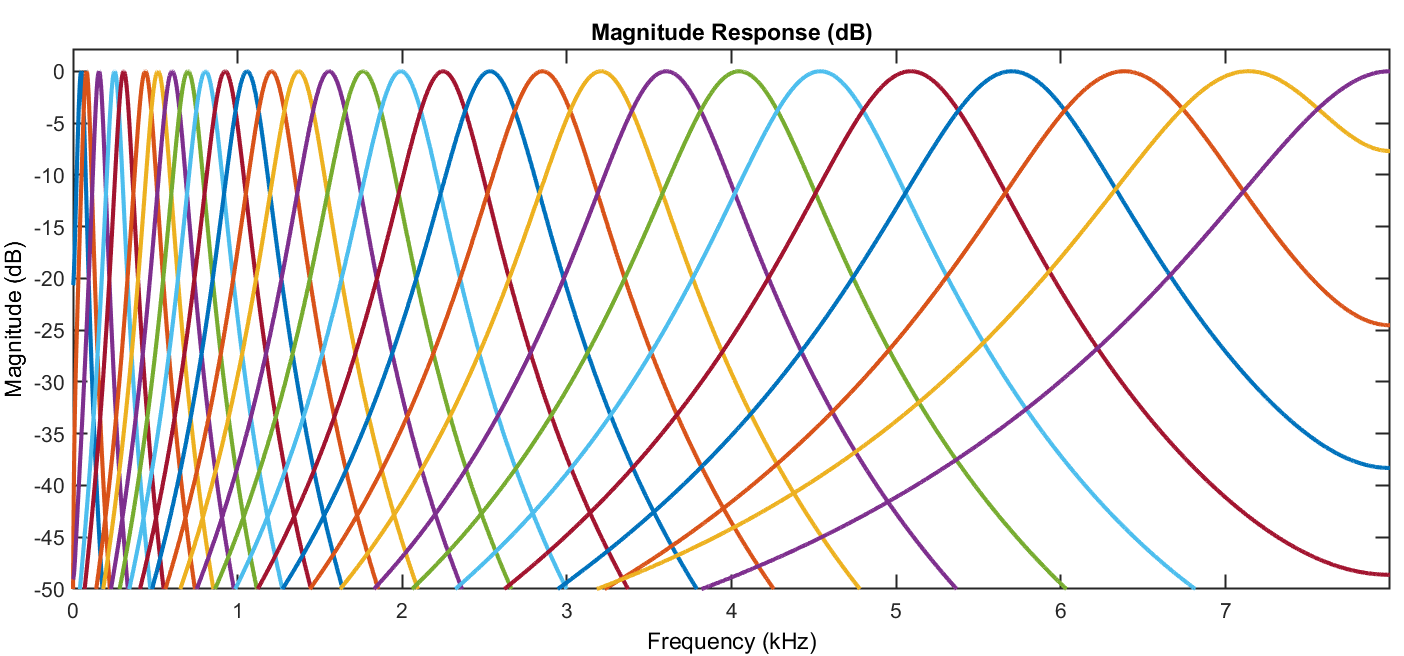}}
\caption{Gammatone filter-bank}
\label{fig:f3}
\end{figure}

\begin{figure}[htb]

%\begin{minipage}[b]{1.0\linewidth}
  \centering
  \centerline{\includegraphics[width=6.5cm]{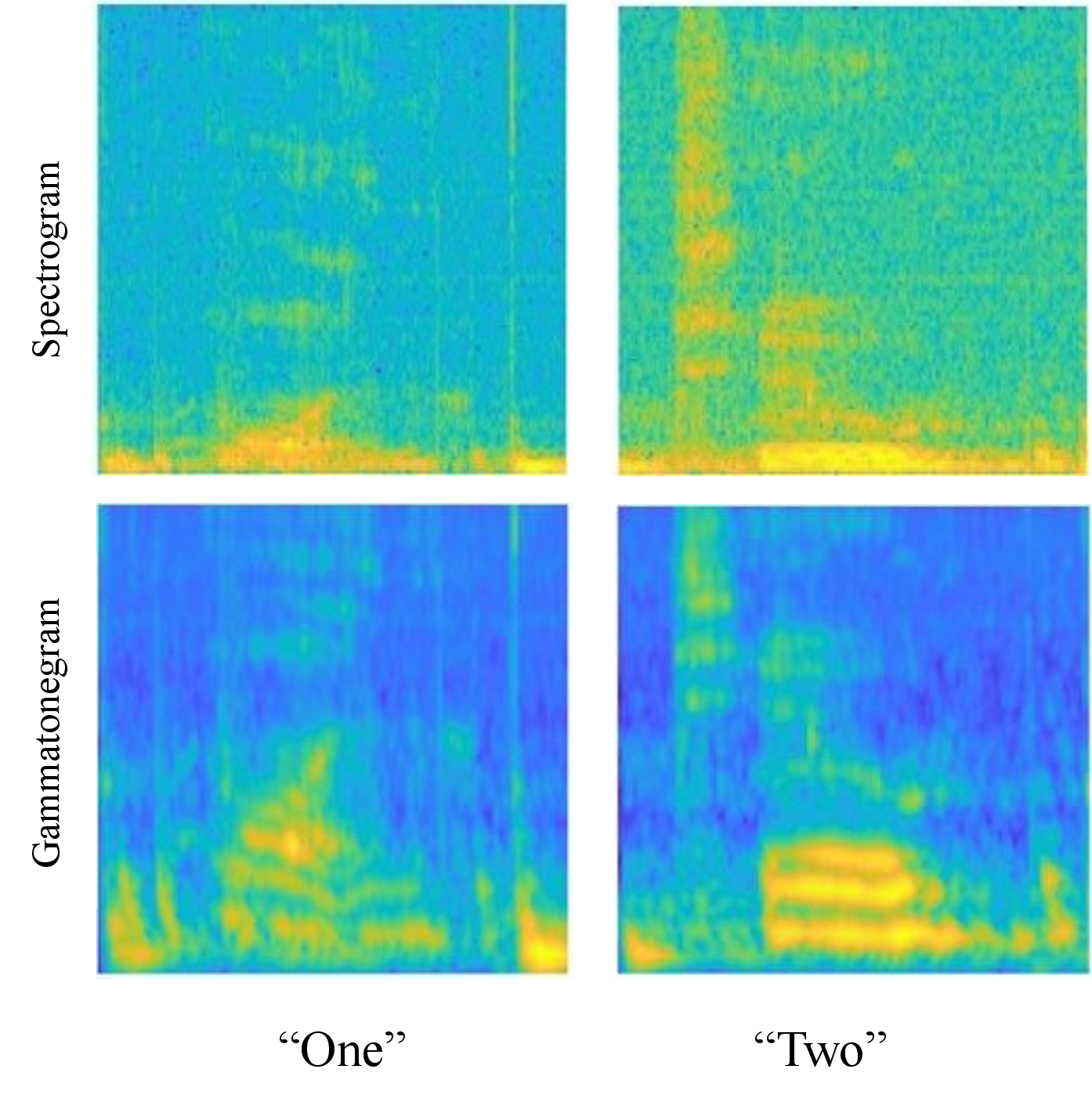}}
\caption{Comparison spectrogram and Gammatonegram representation method in three different utterances}
\label{fig:f4}
\end{figure}

\indent The block diagram presented in Fig. \ref{fig:f2}, illustrates the steps involved in the Gammatonegram extraction. This algorithm has similarities to the spectrogram \cite{rabiner2010theory}, but it offers a more effective representation. The Gammatonegram extraction process begins with pre-emphasis, which involves the utilization of a single-pole filter. This filter compensates for the inherent characteristics of the human speech production system, where high frequencies tend to have lower amplitudes compared to low frequencies. By applying this filter, the energy range in the higher frequencies is increased, resulting in improved intelligibility of the speech. Speech signals are non-static, meaning they cannot be accurately modeled as a combination of sine and cosine functions. Consequently, conventional Fourier transform methods are not suitable for transforming speech signals into the frequency domain. However, within short durations of 20 to 30 milliseconds, speech signals exhibit a more static behavior. To account for this, the speech signal is divided into rectangular frames with a duration of 25 milliseconds.\\
\indent The Gammatonegram extraction process involves applying a hamming window to the rectangular frames before performing the Fourier transform. This windowing technique helps reduce unwanted side lobes that can appear in the transform. To compensate for information loss at the edges, a 10-millisecond overlap is used between frames. The Fourier transform is then applied to the signal, and the amplitude is extracted. Finally, the speech signal is weighted using a Gammatone filter-bank.\\
\indent The Gammatone filter-bank, as depicted in Fig. \ref{fig:f3}, exhibits a high resolution in low frequencies and a low resolution in high frequencies. Multiplying the speech signal with each filter in the filter-bank and summing the outputs of all the filters results in the proposed Gammatonegram representation.\\
\indent The Gammatonegram is represented as an RGB color image, making it suitable for input into a CNN. This type of representation provides higher resolution in low frequencies compared to the traditional spectrogram representation. Fig. \ref{fig:f4} shows an example of these Gammatonegram images compared with the spectrogram to bold the differences. This increased resolution can enhance the discriminative power of different classes. To align with the input layer properties of AlexNet, the final Gammatonegram image is saved in the size of 227x227x3.\\
\subsection{UA Speech Dataset}
\label{ssec:dataset}

\begin{figure}[htb]

%\begin{minipage}[b]{1.0\linewidth}
  \centering
  \centerline{\includegraphics[width=5cm]{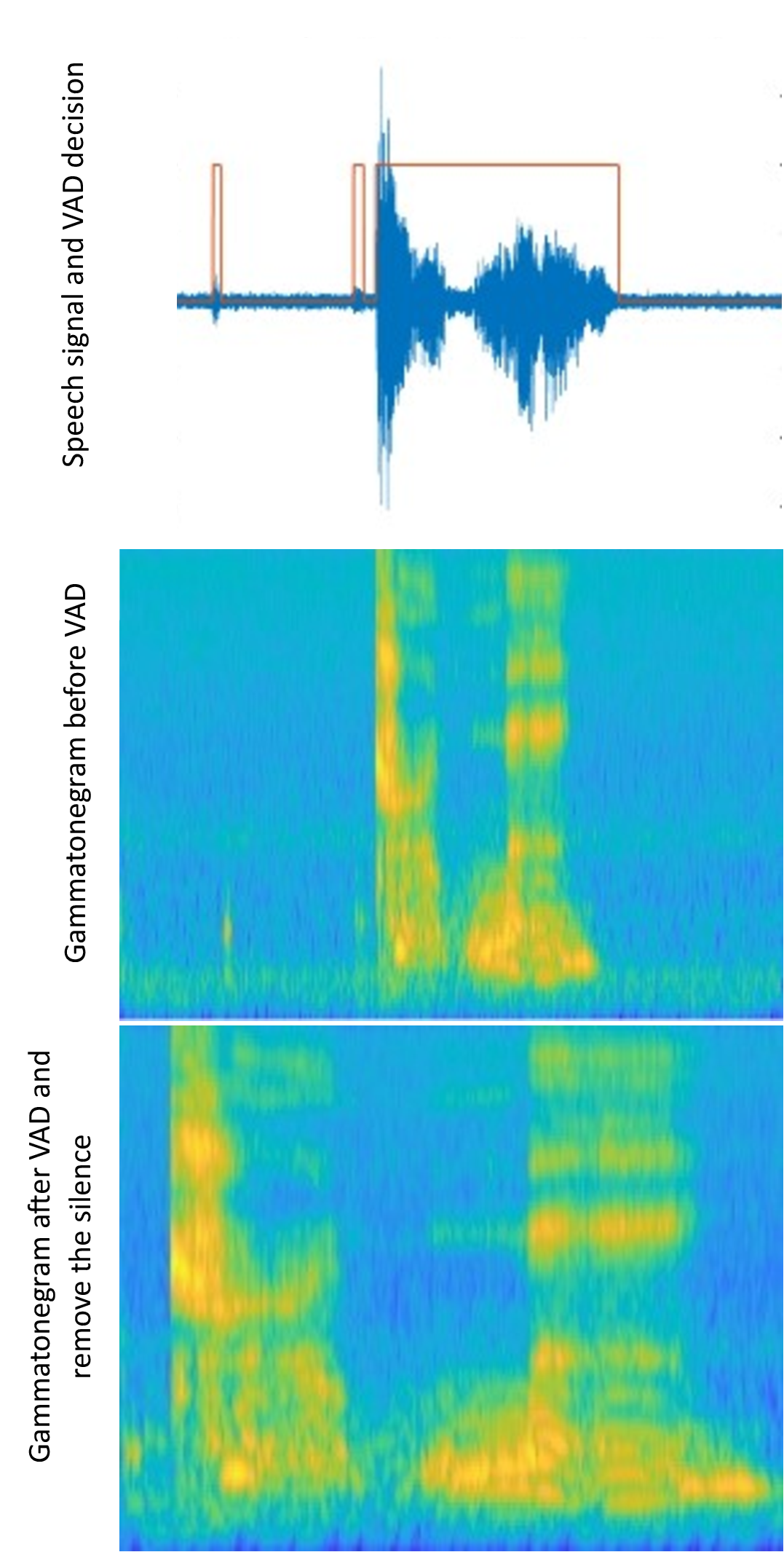}}
\caption{VAD decision and Gammatonegram before and after VAD for a given speech file}
\label{fig:f5}
\end{figure}
\indent A dataset, including 16 dysarthric speakers, has been collected and published by researchers at the University of Illinois \cite{kim2008dysarthric}. These speakers have different severities and speak with varying levels of intelligibility from 2\% to 95\%. The information of the speakers is reported in Table \ref{tab:t1}. This dataset includes 255 isolated dysarthric speech words, consisting of uncommon words, radio alphabet, digits, computer commands, and common worlds. This dataset was collected in three sessions, B1, B2, and B3, with eight microphones. The sampling frequency in this dataset is 16 kHz. It is important to note that this dataset also contains speech files from 12 normal speakers, which were not utilized in this study.
\begin{table}
\caption{Information of the UA speech dataset}
\label{tab:t1}
\centering
\adjustbox{width=8.5cm}{
\begin{tabular}{ccccc}
\textbf{No.} & \textbf{Speaker ID} & \textbf{gender} & \textbf{Age}     & \textbf{Speech Intelligibility} \\ \hline
\textbf{1}   & F02                 & Female          & 30               & 29\%                            \\
\textbf{2}   & F03                 & Female          & 51               & 6\%                             \\
\textbf{3}   & F04                 & Female          & 18               & 62\%                            \\
\textbf{4}   & F05                 & Female          & 22               & 95\%                            \\
\textbf{5}   & M01                 & Male            & \textgreater{}18 & 15\%                            \\
\textbf{6}   & M04                 & Male            & \textgreater{}18 & 2\%                             \\
\textbf{7}   & M05                 & Male            & 21               & 58\%                            \\
\textbf{8}   & M06                 & Male            & 18               & 39\%                            \\
\textbf{9}   & M07                 & Male            & 58               & 28\%                            \\
\textbf{10}  & M08                 & Male            & 28               & 93\%                            \\
\textbf{11}  & M09                 & Male            & 18               & 86\%                            \\
\textbf{12}  & M10                 & Male            & 21               & 93\%                            \\
\textbf{13}  & M11                 & Male            & 48               & 62\%                            \\
\textbf{14}  & M12                 & Male            & 19               & 7.4\%                           \\
\textbf{15}  & M14                 & Male            & 40               & 90.4\%                          \\
\textbf{16}  & M16                 & Male            & \textgreater{}18 & 43\% \\ \hline
\end{tabular}
}
\end{table}
\indent In this study, speech files from 16 dysarthric speakers were used. This subset includes recordings of 30 isolated words, comprising 9 digits, 19 computer commands, and 2 radio alphabets. Each speaker's utterances were saved in eight different files, and these files were found to be almost identical. To ensure reliable evaluations, the K-fold cross-validation method was employed with K=3 because there were three sessions. One session was separated from the other two sessions to avoid excessive similarity between the expressions and prevent any unnatural similarity between the training and testing data. In all experiments, the data from one session was used for training, and two others for testing.
\subsection{Voiced Activity Detector}
\label{ssec:vad}
\indent Silence can have a negative impact on speech processing systems, which is why VAD algorithms are commonly used in such systems. In the case of dysarthric individuals, the inability to pronounce certain syllables, even within a word, often leads to pauses during their speech. Therefore, incorporating VAD can significantly enhance the performance of speech processing systems for these individuals.\\
\indent In our study, we utilize the GMMVAD algorithm \cite{sholokhov2018semi} before representing the speech signal using both the Gammatonegram and spectrogram. This pre-processing step helps to reduce the intra-class variability and can improve the overall efficiency of the system. Fig. \ref{fig:f5} provides an example of the GMMVAD process applied to an audio file, as well as the corresponding Gammatonegram representation before and after applying VAD.\\
\subsection{Evaluation Criteria}
\label{ssec:criteria}
\indent In evaluating the performance of speech recognition systems, various criteria are used. In this study, the Word Recognition Rate (WRR) criterion is employed. WRR calculates the number of isolated words that are correctly recognized compared to the total number of test data.\\
\indent For the speaker identification systems proposed in this work, the network's decision is made based on each audio expression of an isolated word. Therefore, the evaluation involves calculating the number of correct decisions made by the system in comparison to the total number of audio files.\\
\indent In the intelligibility assessment section of the proposed system, each audio file is classified into predetermined categories. The classification is independent of the speaker's identity or speech content. This system's decision is also based on each expression, ensuring that each audio file is evaluated individually.
\section{Experimental Results}
\label{sec:result}
\indent In the experiments, we evaluated the performance of the proposed system based on Gammatonegram representation and the pre-trained CNN in three modes: automatic speech recognition for 30 dysarthric isolated words, dysarthric speaker identification for 16 speakers, speech intelligibility assessment for 2 and 3 class modes, and finally a fully-automated multi-network speech recognition in a cascade architecture.\\
\indent Convolutional neural networks are data hungry, meaning we need lots of data to train a CNN. Transfer learning is a technique to compensate for data shortages in various scenarios. In this work, we first re-train the basic Alexnet with about 40 hours of speech data to recognize dysarthric isolated words in 255 classes. The goal of this work is not to achieve high performance, but we want to give a lot of data to the network so that its feature extraction part can be trained appropriately with Gammatonegram and spectrogram images. This new CNN was used as the pre-trained network to build the systems in all the proposed tasks.\\
\indent Before evaluating our innovative systems, we answer two questions about the proposed method. 1) How is the efficiency of this system compared to a traditional system based on HMM. 2) Does the proposed Gammatonegram perform better than the classical spectrogram. These two questions make up the initial experiments.
\begin{table}
\caption{Overall comparison of the result of the preliminary tests}
\label{tab:t2}
\centering
\adjustbox{width=5cm}{
\begin{tabular}{@{}lc@{}}
\textbf{System}       & \textbf{WRR\%} \\ \hline
HMM-GMM               & 66.23          \\
CNN + Spectrogram   & 86.59          \\
CNN + Gammatonegram & \textbf{91.29} \\ \hline
\end{tabular}
}
\end{table}
\subsection{Initial Experiments}
\label{ssec:initial}
\indent Before the era of deep neural networks, the HMM was one of the most popular methods for speech recognition \cite{sameti2009nevisa,chavan2013overview}. Therefore, we initially evaluated the performance of a traditional HMM-GMM-based ASR system with MFCC feature for dysarthric speech and compared it with the proposed end-to-end systems to highlight the proposed system concept.  In this comparison, the training and test data were completely identical to be a benchmark for measuring performance.\\
\indent In addition to the classification method, we need to pursue the efficiency of the proposed representation method. Therefore, the proposed representation method, i.e., Gammatonegram, should be compared with the conventional representation method, i.e., spectrogram. To this goal, two systems were built separately under the same conditions based on Gammatonegram and spectrogram, in which the number of classes, the amount of training and test data, the network structure, and learning parameters were completely similar.\\
\indent All these three systems were trained for 30 dysarthric isolated words. The system based on HMM-GMM has three states and four Gaussians in each state. The MFCC features, energy, and first and second-order derivatives have been extracted from the audio signal, totaling 39 features per frame. These parameters have been chosen based on lots of experiments. It should be noted that the proposed HMM system was implemented using Murphy Toolbox \cite{murphy1998hidden}. However, we trained the proposed CNN network using the introduced pre-trained network for Gammatonegram and spectrogram separately.\\
\indent Based on results in Table \ref{tab:t2}, the HMM-based system achieved 66.23\% overall WRR, which is poor performance compared to the other two systems. The CNN-based systems show an acceptable performance despite the insufficient training data. Meanwhile, the Gammatonegram representation system shows better results and reaches a 91.29\% WRR. These results verify that the proposed Gammatonegram method for representation and CNN for end-to-end classification are the right choices for dysarthric speech processing.
\subsection{Automatic Speech Recognition}
\label{ssec:ASR}
\indent For disabled people, having a smart home system based on artificial intelligence can be helpful. One of the best ways to interact with this system is through speech signals. In this case, by checking the contents of the speech file, the ASR system tries to identify the command word. In this system, the information related to speech content is important, not the speaker's identity. Therefore, this system generally operates in speaker-dependent (SD) and Speaker-Independent (SI) modes. In the SD mode, the speakers' identity in the training and test phases are the same and the network adapts to these speakers' information. In this case, the system is more efficient because it is familiar with the parameters related to the speakers. However, In the SI mode, there is no information about test speakers in the training phase. The performance of ASR systems usually decreases in SI mode because the information related to the test speakers affects their performance.\\
\indent In this section, proposed dysarthric ASR systems are evaluated in both modes. A unique CNN was trained for all the speakers in the SD mode. In SI mode, there is a specific ASR system for each speaker. To evaluate the proposed ASR systems, 51 models have been trained for all modes and folds. To create these systems in SI mode, each test speaker's speech files were left out, and the system was trained using the speech of other speakers. The simulation was repeated for all 16 speakers, and a specific SI network was trained for each speaker. In Table \ref{tab:t3}, the results of the proposed ASR systems are reported.\\
\begin{table}
\caption{Results of automatic speech recognition systems in SD and SI scenarios}
\label{tab:t3}
\centering
\adjustbox{width=6cm}{
\begin{tabular}{@{}lcc@{}}
\textbf{Spkr} & \textbf{WRR in SD (\%)} & \textbf{WRR in SI (\%)} \\ \midrule
F02           & 98.19                        & 86.63                        \\
F03           & 80.18                        & 63.82                        \\
F04           & 95.59                        & 93.18                        \\
F05           & 97.93                        & 95.28                        \\
M01           & 88.28                        & 83.62                        \\
M04           & 68.06                        & 51.67                        \\
M05           & 92.63                        & 90.95                        \\
M06           & 94.16                        & 78.81                        \\
M07           & 85.71                        & 85.70                         \\
M08           & 98.85                        & 95.71                        \\
M09           & 98.62                        & 97.57                        \\
M10           & 98.85                        & 97.14                        \\
M11           & 93.01                        & 88.33                        \\
M12           & 78.49                        & 61.87                        \\
M14           & 96.43                        & 89.93                        \\
M16           & 95.70                         & 91.83                        \\ \midrule
\textbf{Mean} & \textbf{91.29}               & \textbf{84.50}               \\ \bottomrule
\end{tabular}
}
\end{table}
\indent In these experiments, the CNNs were trained with batch size 32, which was the best choice based on our computational resources, and based on several experiments with different amounts for epoch numbers, we found that 20 was the best choice. The ASR system in the SD mode achieved an average WRR of 91.29\%, which is about 6.5\% better than the SI mode with 84.50\% WRR. In addition, by analyzing the results for each speaker, it can be found the system has its lowest performance for speakers with high severity. In detail, the system's performance for M04 and F03 was worse. It was because of the very low intelligibility of their speech that the characteristic features of speech were strongly destroyed. This was because of less control in muscles participating in the speech production mechanism. However, the proposed system learned the normal speech features properly and performed well for high-intelligibility speech, such as speech files belonging to F05, M08, and M09. Results showed that our proposed Gammatonegram method, in cooperation with the end-to-end ASR system, has acceptable performance for dysarthric speech because of the high potential to represent the speech contents.
\subsection{Automatic Speaker Identification}
\label{ssec:ASI}
\indent In scenarios like smart homes, the voice key is beneficial for disabled individuals because in cases such as locking the door or permission to access control, speaker identification can allow the disabled person to gain access. Therefore, designing an efficient speaker identification system can be helpful. The proposed systems were evaluated in two Text-Dependent (TD) and text-independent (TI) modes. We trained a CNN for each one of the scenarios and these CNNs were trained with about 5 minutes of speech for each speaker. The UA speech dataset consists of 16 dysarthric speakers, so the output layer has 16 classes, each representing one of the speakers.\\
\indent The texts expressed in the test and training phases are the same in the TD mode. In other words, the dysarthric person has to repeat a specific password in both stages. The system was tested with two sessions' data of the UA dataset. However, the speech contents used for training and testing in TI mode are different. In other words, in this case, a person can use any word as a voice password, and the system recognizes the person's identity with different speech content outside of the training data. For the test of the TI system, the CW1 to CW50 words of the UA dataset, which had not been used in the training phase, were used. The systems were trained with batch size 32, and 30 iterations based on several evaluations to find the best parameter measure. The results obtained from both modes were reported in Table \ref{tab:t4}. The performance of the systems reached 87.74\% accuracy in TD mode and 80.70\% in TI mode. In speaker identification systems, like ASR systems, speakers with low speech intelligibility rates, such as F03 and M12, are the reduction agents in the recognition rate. This performance was acquired in low training data conditions and depicted that Gammatonegram contains speaker-specific features.
\begin{table}
\caption{Results of speaker identification systems in text-dependent and text-independent modes}
\label{tab:t4}
\centering
\adjustbox{width=7.3cm}{
\begin{tabular}{@{}lcc@{}}
\textbf{Spkr} & \textbf{Text-Dependent   (\%)} & \textbf{Text-Independent   (\%)} \\ \midrule
F02           & 95.10               & 81.50               \\
F03           & 89.89              & 76.50               \\
F04           & 95.34              & 91.75              \\
F05           & 98.38              & 88.03              \\
M01           & 94.56              & 90.90               \\
M04           & 84.19              & 79.47              \\
M05           & 75.34              & 58.39              \\
M06           & 89.71              & 66.76              \\
M07           & 88.47              & 88.20               \\
M08           & 64.51              & 65.47              \\
M09           & 91.24              & 79.57              \\
M10           & 80.41              & 64.09              \\
M11           & 86.82              & 80.99              \\
M14           & 80.95              & 86.71              \\
M16           & 90.05              & 93.34              \\ \midrule
\textbf{Mean} & \textbf{87.74}   & \textbf{80.70}  \\ \bottomrule
\end{tabular}
}
\end{table}
\subsection{Cascade System For Multi-Network ASR}
\label{ssec:Cascade}
\begin{figure*}
  \includegraphics[width=\textwidth]{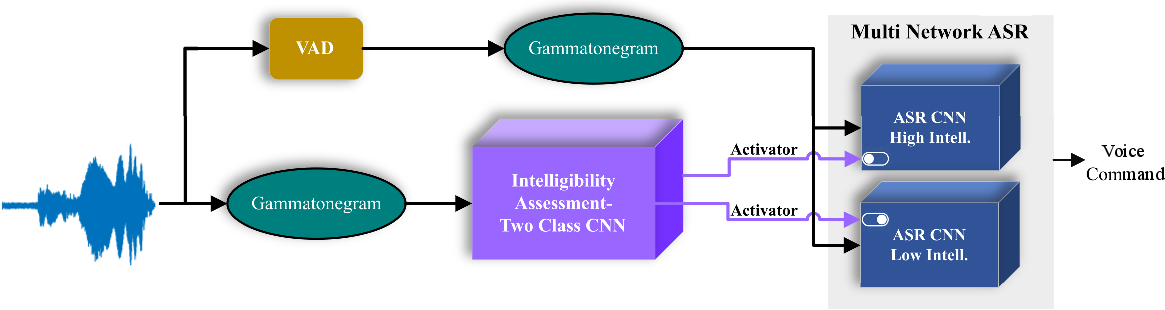}
%  \vspace{2.0cm}

%\end{minipage}

\caption{Structure of multi-network speech recognition system in cascade architecture with two-class automatic intelligibility assessment}
\label{fig:f6}
\end{figure*}

\indent In previous dysarthric speech processing studies, multi-network architectures have been utilized \cite{farhadipour2018dysarthric,shahamiri2014real}. However, none of these studies have automated the process of assigning audio files to the appropriate network. Instead, individuals with dysarthria were required to manually determine which network or category their speech belonged to. In our proposed multi-network cascade architecture, we introduce an intelligibility assessment system that automatically activates one of the multi-networks for ASR. This architecture, depicted in Fig. \ref{fig:f6}, consists of two main steps. According to this figure, in the first step, the intelligibility assessment system classifies incoming speech into two categories: high intelligibility and low intelligibility. In the second step, we trained two ASR systems for each intelligibility category.\\
\indent Automatic process the disabled people's speech to determine their speech intelligibility level is effective for many purposes. For instance, automatically diagnose the disease severity and the growth process of disability by periodically checking their speech. Moreover, the automatic intelligibility assessment can improve the efficiency of ASR and speaker identification systems in multi-network scenarios. In this scenario, we trained several parallel networks for ASR. The dysarthric speakers expressed speech commands without knowledge of the multi-network structure or even the severity level of their disability. Automatic intelligibility assessment examines the person's speech and assigns it to the corresponding network according to the intelligibility level.\\
\indent  For this purpose, different categories were made according to the intelligibility percentage. In this study, according to the efficiency of the system and the amount of available data, the speakers are divided into three-class and two-class modes based on the intelligibility level, and two separate networks were trained to recognize the intelligibility. The interesting point in this scenario is that the speech of dysarthric individuals is sometimes accompanied by unusual silence, especially for explosive phonemes in the middle of a word. This phenomenon can play an essential role in determining the intelligibility level of a dysarthric person's speech. For this reason, intelligibility assessment systems were trained and evaluated without VAD. In this case, CNN networks were trained using batch size 32, and 20 iterations.\\
\begin{table*}
\caption{Results of two automatic intelligibility assessments and results of two proposed architecture of cascade speech recognition systems}
\label{tab:t5}
\small
\centering
\begin{tabular}{p{3cm}@{}lcclllllccc@{}}
                                & \multicolumn{2}{c}{\textbf{Intelligibility (\%)}} &  &  &  &                     &                                 & \multicolumn{1}{l}{}                                                                                    & \multicolumn{2}{c}{\textbf{Cascade ASR in SD (\%)}}           \\ \cmidrule(lr){2-3} \cmidrule(l){10-11}
\multirow{-2}{*}{\textbf{Spkr}} & \textbf{3-Class}        & \textbf{2-Class}        &  &  &  &                     & \multirow{-2}{*}{\textbf{Spkr}} & \multicolumn{1}{l}{\multirow{-2}{*}{\textbf{Severity}}}                                                 & \textbf{3-Class}              & \textbf{2-Class}              \\ \cmidrule(r){1-3} \cmidrule(l){8-11}
F02                             & 97.06                   & 98.21                   &  &  &  &                     & \cellcolor[HTML]{A6A6A6}F02     & \cellcolor[HTML]{A6A6A6}                                                                                & \cellcolor[HTML]{A6A6A6}92.99 & \cellcolor[HTML]{A6A6A6}94.13 \\
F03                             & 100                     & 100                     &  &  &  &                     & \cellcolor[HTML]{A6A6A6}F03     & \cellcolor[HTML]{A6A6A6}                                                                                & \cellcolor[HTML]{A6A6A6}71.03 & \cellcolor[HTML]{A6A6A6}81.89 \\
M01                             & 87.23                   & 93.31                   &  &  &  &                     & \cellcolor[HTML]{A6A6A6}M01     & \cellcolor[HTML]{A6A6A6}                                                                                & \cellcolor[HTML]{A6A6A6}75.99 & \cellcolor[HTML]{A6A6A6}83.28 \\
M04                             & 94.89                   & 99.33                   &  &  &  &                     & \cellcolor[HTML]{A6A6A6}M04     & \cellcolor[HTML]{A6A6A6}                                                                                & \cellcolor[HTML]{A6A6A6}71.11 & \cellcolor[HTML]{A6A6A6}81.56 \\
M07                             & 89.05                   & 99.76                   &  &  &  &                     & \cellcolor[HTML]{A6A6A6}M07     & \cellcolor[HTML]{A6A6A6}                                                                                & \cellcolor[HTML]{A6A6A6}88.1  & \cellcolor[HTML]{A6A6A6}93.57 \\
M12                             & 98.33                   & 100                     &  &  &  &                     & \cellcolor[HTML]{A6A6A6}M12     & \multirow{-6}{*}{\cellcolor[HTML]{A6A6A6}\begin{tabular}[c]{@{}c@{}}High\\    \\ 2\%-37\%\end{tabular}} & \cellcolor[HTML]{A6A6A6}73.7  & \cellcolor[HTML]{A6A6A6}86.11 \\
F04                             & 79.8                    & 92.72                   &  &  &  &                     & \cellcolor[HTML]{BFBFBF}F04     & \cellcolor[HTML]{BFBFBF}                                                                                & \cellcolor[HTML]{BFBFBF}92.72 & \cellcolor[HTML]{A6A6A6}95.36 \\
M05                             & 89.37                   & 93.81                   &  &  &  &                     & \cellcolor[HTML]{BFBFBF}M05     & \cellcolor[HTML]{BFBFBF}                                                                                & \cellcolor[HTML]{BFBFBF}92.06 & \cellcolor[HTML]{A6A6A6}94.13 \\
M06                             & 97.04                   & 94.78                   &  &  &  &                     & \cellcolor[HTML]{BFBFBF}M06     & \cellcolor[HTML]{BFBFBF}                                                                                & \cellcolor[HTML]{BFBFBF}89.91 & \cellcolor[HTML]{A6A6A6}95.13 \\
M11                             & 74.07                   & 89.94                   &  &  &  &                     & \cellcolor[HTML]{BFBFBF}M11     & \cellcolor[HTML]{BFBFBF}                                                                                & \cellcolor[HTML]{BFBFBF}88.52 & \cellcolor[HTML]{A6A6A6}93.33 \\
M16                             & 86.3                    & 92.22                   &  &  &  &                     & \cellcolor[HTML]{BFBFBF}M16     & \multirow{-5}{*}{\cellcolor[HTML]{BFBFBF}\begin{tabular}[c]{@{}c@{}}Mid\\    \\ 35\%-62\%\end{tabular}} & \cellcolor[HTML]{BFBFBF}91.85 & \cellcolor[HTML]{A6A6A6}94.07 \\
F05                             & 97.62                   & 98.1                    &  &  &  &                     & \cellcolor[HTML]{D9D9D9}F05     & \cellcolor[HTML]{D9D9D9}                                                                                & \cellcolor[HTML]{D9D9D9}97.94 & \cellcolor[HTML]{D9D9D9}98.1  \\
M08                             & 98.73                   & 98.41                   &  &  &  &                     & \cellcolor[HTML]{D9D9D9}M08     & \cellcolor[HTML]{D9D9D9}                                                                                & \cellcolor[HTML]{D9D9D9}96.83 & \cellcolor[HTML]{D9D9D9}96.83 \\
M09                             & 98.41                   & 97.94                   &  &  &  &                     & \cellcolor[HTML]{D9D9D9}M09     & \cellcolor[HTML]{D9D9D9}                                                                                & \cellcolor[HTML]{D9D9D9}95.56 & \cellcolor[HTML]{D9D9D9}95.56 \\
M10                             & 98.09                   & 97.77                   &  &  &  &                     & \cellcolor[HTML]{D9D9D9}M10     & \cellcolor[HTML]{D9D9D9}                                                                                & \cellcolor[HTML]{D9D9D9}98.41 & \cellcolor[HTML]{D9D9D9}98.73 \\
M14                             & 97.89                   & 97.24                   &  &  &  &                     & \cellcolor[HTML]{D9D9D9}M14     & \multirow{-5}{*}{\cellcolor[HTML]{D9D9D9}\begin{tabular}[c]{@{}c@{}}Low\\    \\ 63\%-95\%\end{tabular}} & \cellcolor[HTML]{D9D9D9}95.62 & \cellcolor[HTML]{D9D9D9}95.13 \\ \cmidrule(r){1-3} \cmidrule(l){8-11}
\textbf{Mean}                   & \textbf{92.74}          & \textbf{96.47}          &  &  &  & \multirow{-19}{*}{} & \textbf{Mean}                   & \multicolumn{1}{l}{\textbf{}}                                                                           & \textbf{88.27}                & \textbf{92.30}                \\ \cmidrule(r){1-8} \cmidrule(l){8-11}
\end{tabular}
\end{table*}
\indent  Table \ref{tab:t5} reports the results of three- and two-class networks. In the three-class mode, speakers were classified into three categories: high, mid, and low, whose intelligibility range in each class is shown in Table \ref{tab:t5}. In the two-class mode, the high and mid categories were combined because we realized a high correlation between data for these two classes. However, the low severity category remains unchanged. These two systems were trained in SD mode, in which one session of the dataset was used for training and two others were acquired for testing. According to the results, the performance has improved in the two-class mode, so the average intelligibility recognition accuracy using CNN and Gammatonegram in the two and three classes have reached 96.47\% and 92.74\%, respectively.\\
\indent Part 2 of Table \ref{tab:t5} provides the results of the multi-network ASR in cascade structure with the intelligibility assessment system. The results are reported in two and three-class modes. According to these results, the performance of the speech recognition system in the dual-network improved compared to the single-network mode and reached 92.3\% WRR in the SD mode. This achievement was because each network focuses on close-range speech intelligibility or less intra-class variation.
\section{Comparative Analysis of Proposed Systems}
\label{sec:comp}
\begin{figure*}[t]
\centering
  \includegraphics[width=15cm]{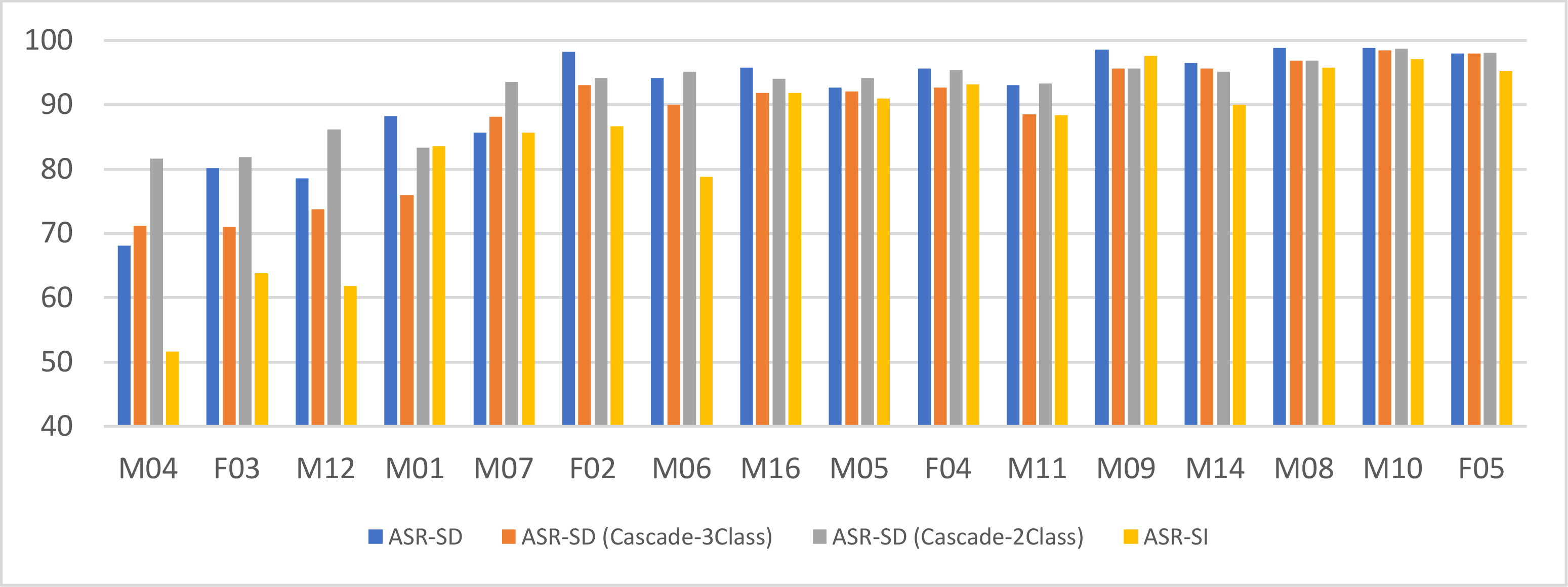}
\caption{Comparison of the performance of dysarthric speech recognition systems in single and multi-network scenarios of different speakers}
\label{fig:f7}
\centering
\includegraphics[width=15cm]{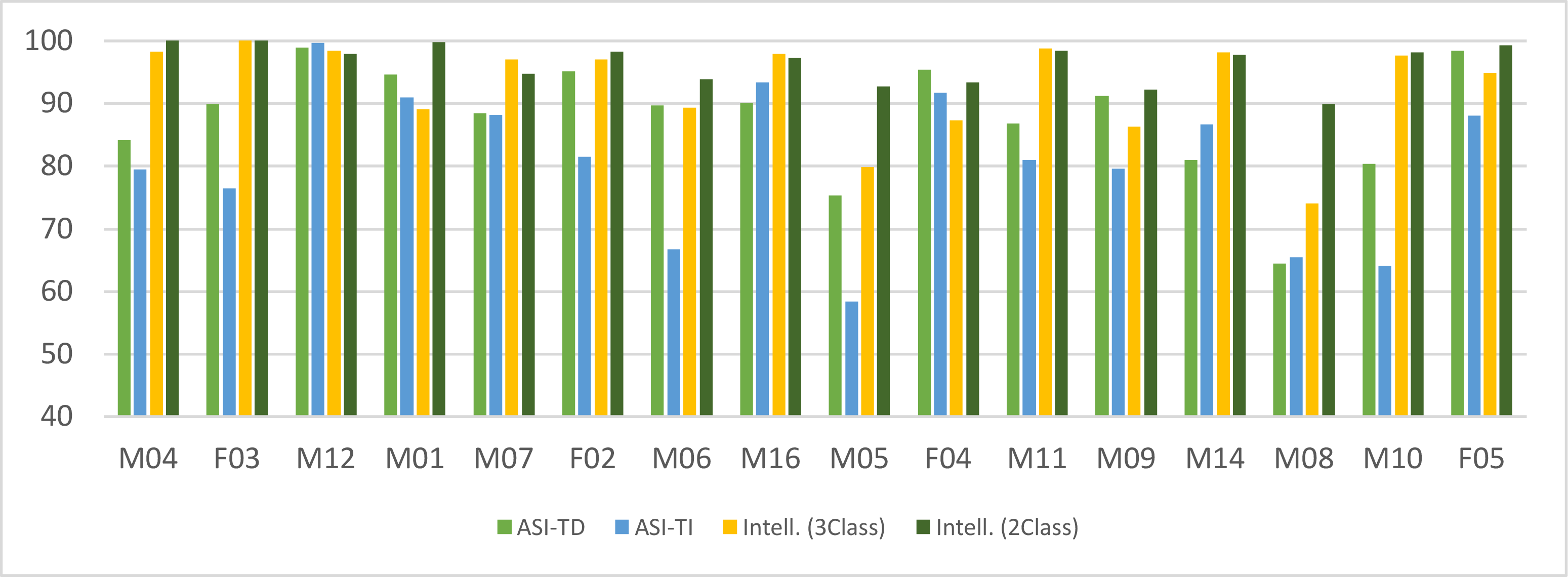}
\caption{Comparison of the performance of speaker recognition and intelligibility assessment systems of different speakers}
\label{fig:f8}
\end{figure*}

\indent The performance of proposed ASR systems in different modes is shown in Fig. \ref{fig:f7} so that it can be analyzed more efficiently for each speaker. In this chart, the speakers are sorted based on dysarthric severity from the highest to the lowest, as reported in the dataset. In the single network, both in the SD and SI modes, the performance was consistently lower than the average for the first five speakers who had the highest severity of dysarthria. This can be attributed to the variability and instability of the dysarthric speech signal in individuals with high severity, leading to system errors. Conversely, the recognition rate for the low-severity group was higher than the average, as their speech parameters closely resembled normal speech with a predictable form and minimal diversity between the test and training data.\\
\indent The proposed multi-network ASR system, particularly in the two-class mode, demonstrated a significant improvement in performance for the high-severity group. This improvement was achieved by designing a network that specifically focused on the parameters of the high-severity group, which differed significantly from the other two groups. Consequently, this network efficiently learned the parameters of the high-severity group's speech.\\
\indent Figure \ref{fig:f8} illustrates the performance of the speaker identification and intelligibility assessment systems. Based on the results, there seems to be a low correlation between the speaker identification system's performance and the severity of dysarthria in comparison with the ASR system. However, Gammatonegram performed well in the intelligibility assessment task, validating our hypothesis that using Gammatonegram without VAD is effective, as the system's efficiency was deemed acceptable. Leveraging the achievements and performance of Gammatonegram, we subsequently designed our multi-network fully automated ASR system based on the intelligibility assessment approach.\\
\indent The performance of Gammatonegram in the ASR task reached a WRR of 84.50\% in the SI mode and 91.29\% in the SD mode. In the speaker identification task, our proposed system achieved recognition rates of 80.70\% and 87.74\% in the TI and TD modes, respectively. Moreover, Gammatonegram performed well in the intelligibility assessment task, with average recognition rates of 92.74\% for the three-class mode and 96.47\% for the two-class mode. Finally, the proposed cascade ASR system achieved 92.3\% WRR. A detailed comparison with previous works based on their respective tasks is provided in Table \ref{tab:t6} to better understand our achievements.\\
\begin{table*}
\caption{Results of two automatic intelligibility assessments and results of two proposed architecture of cascade speech recognition systems}
\label{tab:t6}
\small
\centering
\begin{tabular}{clll}
\multicolumn{1}{l}{\textbf{Task}}                 & \textbf{Reference}                              & \textbf{WRR (\%)} & \textbf{Method}            \\ \hline
\multicolumn{1}{c|}{\multirow{7}{*}{ASR}}         & \cite{rathod2023transfer}      & 59.78                  & Bi-LSTM                     \\
\multicolumn{1}{c|}{}                             & \cite{shahamiri2021speech}     & 64.71                  & Voicegram                  \\
\multicolumn{1}{c|}{}                             & \cite{shahamiri2023dysarthric} & 67.00                  & Deep Transformers          \\
\multicolumn{1}{c|}{}                             & \cite{liu2021recent}           & 74.79                  & Visual Features            \\
\multicolumn{1}{c|}{}                             & \cite{almadhor2023e2e}         & 79.28                  & E2E                        \\
\multicolumn{1}{c|}{}                             & \cite{yu2023multi}             & 86.50                  & AV-HuBERT                  \\
\multicolumn{1}{c|}{}                             & \textbf{Proposed ASR}                           & \textbf{92.30}         & \textbf{Cascade system}    \\ \hline
\multicolumn{1}{c|}{\multirow{3}{*}{Spkr Ident.}} & \cite{salim2023constant}       & 84.93                  & MFCC+ivector               \\
\multicolumn{1}{c|}{}                             & \cite{salim2022automatic}      & 87.34                  & xvector                    \\
\multicolumn{1}{c|}{}                             & \textbf{Proposed System}                        & \textbf{87.74}         & \textbf{E2E+Gammatonegram} \\ \hline
\multicolumn{1}{c|}{\multirow{4}{*}{Intell. A.}}  & \cite{nikhil2023few}           & 85.00                  & Transformer                \\
\multicolumn{1}{c|}{}                             & \cite{joshy2022automated}       & 93.97                  & DNN+Prosody Feature        \\
\multicolumn{1}{c|}{}                             & \cite{al2021classification}    & 95.80                  & Acoustic Feature           \\
\multicolumn{1}{c|}{}                             & \textbf{Proposed System}                        & \textbf{96.47}         & \textbf{E2E+Gammatonegram} \\ \hline
\end{tabular}
\end{table*}
\indent Based on the results and the comparison with previous studies, it is evident that the Gammatongram representation effectively captures the speech characteristics of individuals with dysarthria. Additionally, the utilization of a two-dimensional convolutional network demonstrates strong performance. Notably, the proposed Cascade network introduces a novel approach to speech recognition for dysarthric individuals, allowing for the seamless integration of multi-network ASR in a fully automated manner.

\section{Conclusion}
\label{sec:conc}
\indent In this work, we introduced Gammatonegram as an adequate representation method and utilized transfer learning to build end-to-end dysarthric speech processing systems based on CNNs. The introduced systems have been evaluated in three tasks: speech command recognition, speaker identification, and intelligibility assessment. Before considering the proposed methods, we compare the performance of a traditional ASR system based on HMM-GMM with our proposed end-to-end system based on Gammatonegram representation. Results depicted that the proposed system outperformed in an ASR scenario with a significant interval. Another comparison has been made to verify our proposed Gammatonegram with a traditional spectrogram as a popular method for representing speech signals as an image in a similar situation. Results verified all subsequent simulations using the proposed method.\\
\indent The proposed systems utilized the UA dysarthric speech dataset and employed the GMMVAD algorithm for silence removal. The widely recognized Alexnet was chosen as the initial network and then retrained using 255 audio commands. This retraining process focused on training the first part of the network, which was responsible for feature extraction, with a substantial number of Gammatonegram images. This pre-trained network was then employed to model all scenarios using the transfer learning technique. In each Folds evaluation, Only one session was utilized for system training, while two others were used for system evaluation.\\
\indent In the first task, speech recognition systems were designed and evaluated in speaker-dependent and speaker-independent modes based on the Gammatonegram representation. The results demonstrated that the proposed system achieved acceptable performance. It was observed from the results that the progression of the disease in individuals had an inverse relationship with the efficiency of the speech recognition system for their speech. In other words, the system was less efficient for the speech from individuals with more severe diseases.\\
\indent Moving on to the second task, the objective was to recognize the identity from the speech signal. Two scenarios, namely text-independent and text-dependent, were evaluated. The efficiency of the systems in this task revealed that the Gammatonegram representation contains valuable information about the speaker, which enables the system to recognize their identity.\\
\indent The third task focused on intelligibility assessment, conducted in two- and three-class scenarios. Since silence within each word also plays a crucial role in speech intelligibility, the VAD was not employed in this task. The results indicated that speech intelligibility assessment performs better in the two-class mode and can be used as a complementary tool for new tasks, such as multi-network speech recognition.\\
\indent Lastly, we developed an automatic multi-network system for ASR. This system automatically assigned input speech utterances to corresponding speech recognition networks based on the intelligibility percentage. Using a cascade architecture and a two-class speech recognition approach, the system achieved a WRR of 92.3\%, indicating an improvement compared to the single-network mode.\\
\indent Future studies could further improve the results by implementing a cascade approach for speaker identification tasks. In addition, incorporating data augmentation techniques could be beneficial. By adding different types of noises and music to the speech files, the system can be trained to be more robust and adaptable to real-world scenarios. The source code of this paper is available \footnote[1]{\url{https://github.com/areffarhadi/Gammatonegram_CNN_Dysarthric_speech}}.
\\

\textbf{Declarations}\\
\textbf{Ethical Approval: }
This paper reflects the authors' own research and analysis truthfully and completely and is not currently being considered for publication elsewhere.
 
\textbf{Competing interests: }
The authors declare that they have no known competing financial interests or personal relationships that could have appeared to influence the work reported in this paper.
 
\textbf{Authors' contributions: }  
In preparing this paper, all the authors' shares of contributions were equal. 
 
\textbf{Funding: }
The authors did not receive support from any organizations or sources for the submitted work.
 
\textbf{Availability of data and materials: }  
The source code of this paper is available. Moreover, The UASpeech dataset is available freely.

\vfill\pagebreak

\bibliographystyle{IEEEbib}
\bibliography{dys_Template}

\end{document}